\documentclass[prl,twocolumn,superscriptaddress,nobibnotes,letterpaper]{revtex4}

\usepackage{times}
\usepackage[pdftex]{graphicx}
\usepackage[pdftex]{epsfig}
\usepackage{amsmath}
\usepackage{pslatex}
\usepackage{amssymb}
\usepackage{amsfonts}
\usepackage{color}
\usepackage{wrapfig}
\usepackage{eucal}
\usepackage{hhline}
\usepackage{threeparttable}
\usepackage{supertabular}
\usepackage{multirow}
\usepackage{tabularx}

\newcommand{\nbar}{\langle n \rangle}

\newcommand{\avg}[1]{\mbox{$\langle#1\rangle$}}

\newcolumntype{Y}{>{\centering\arraybackslash}X}

\def\be{\begin{equation}}
\def\ee{\end{equation}}
\def\bea{\begin{eqnarray}}
\def\eea{\end{eqnarray}}
\newcommand{\opdagger}[2]{\mbox{$\hat{#1}_{#2}^{\dagger}$}}
\newcommand{\op}[2]{\mbox{$\hat{#1}_{#2}$}}

\begin{document}

\pagenumbering{arabic}

\title{Measurement of the quantum zero-point motion of a nanomechanical resonator}

\author{Amir H. Safavi-Naeini}
\email{safavi@caltech.edu}
\author{Jasper Chan}
\author{Jeff T. Hill}
\author{T. P. Mayer Alegre}
\author{Alex Krause}
%\affiliation{Thomas J. Watson, Sr., Laboratory of Applied Physics, California Institute of Technology, Pasadena, CA 91125}
%\author{Simon Gr\"oblacher}
%\affiliation{Thomas J. Watson, Sr., Laboratory of Applied Physics, California Institute of Technology, Pasadena, CA 91125}
%\affiliation{Vienna Center for Quantum Science and Technology (VCQ), Faculty of Physics, University of Vienna, Boltzmanngasse 5, %A-1090 Vienna, Austria}
%\author{Markus Aspelmeyer}
%\affiliation{Vienna Center for Quantum Science and Technology (VCQ), Faculty of Physics, University of Vienna, Boltzmanngasse 5, %A-1090 Vienna, Austria}
\author{Oskar Painter}
\email{opainter@caltech.edu}
\homepage{http://copilot.caltech.edu}
\affiliation{Thomas J. Watson, Sr., Laboratory of Applied Physics, California Institute of Technology, Pasadena, CA 91125}

\date{\today}

\begin{abstract}
We present optical sideband spectroscopy measurements of a mesoscopic mechanical oscillator cooled near its quantum ground state.  The mechanical oscillator, corresponding to a $3.99~\text{GHz}$ acoustic mode of a patterned silicon nanobeam, is coupled via radiation pressure to a pair of co-localized $200$~THz optical modes.  The mechanical mode is cooled close to its quantum ground state from a bath temperature of $T_b \approx 18$~K using radiation pressure back-action stemming from the optical pumping of one of the optical cavity resonances. An optical probe beam, resonant with the second optical cavity resonance, is used to transduce the mechanical motion and determine the phonon occupancy of the mechanical mode.  Measurement of the asymmetry between up-converted and down-converted photons of the probe beam yields directly the displacement noise power associated with the quantum zero-point motion of the mechanical oscillator, and provides an absolute calibration of the average phonon occupancy of the mechanical mode.  
\end{abstract}

\maketitle

%% Introduction

The Heisenberg uncertainty principle, one of the fundamental consequences of quantum theory, restricts the certainty with which the position and momentum of an object may be simultaneously known and defined.  Consequently, an object confined to a local potential possesses a non-zero ground state energy associated with random quantum fluctuations of its position.  For mechanical systems of our daily experience, this so-called quantum zero-point motion is masked by thermal noise resulting from interaction with the environment. To observe quantum motion, the dual and antagonistic feats of isolating a mechanical system from its environment whilst measuring its position must be achieved~\cite{ref:Braginsky1977,Caves1980a,Clerk2010}. Recently, coupled electro- and opto-mechanical systems have been cooled into their mechanical quantum ground state by a combination of cryogenic pre-cooling and radiation pressure back-action~\cite{Teufel2011b,Chan2011}.  In this work, we extend these results by laser cooling the acoustic mode of a silicon optomechanical crystal resonator to near its quantum ground state (occupancy $2.6\pm0.2$ phonons) while simultaneously monitoring its motion with a probe laser resonant with a secondary optical mode of the cavity.  Operating in the resolved sideband regime, the spectral selectivity of the secondary cavity mode is used to separate, measure, and compare the intensity of the blue- (anti-Stokes) and red-shifted  (Stokes) sidebands created by the mechanical resonator's motion. The observed asymmetry in the probe sideband amplitudes provides a direct measure of the displacement noise power associated with quantum zero-point fluctuations and allows for an intrinsic calibration of the phonon occupation number of the nanomechanical resonator.

Experiments with trapped atomic ions and neutral atoms~\cite{Diedrich1989,Jessen1992,Monroe1995}, dating back several decades, utilized techniques such as resolved sideband laser cooling and motional sideband absorption and fluorescence spectroscopy to cool and measure a single trapped particle in its vibrational quantum ground state.  These experiments generated significant interest in the coherent control of motion and the quantum optics of trapped atoms and ions\cite{Blockley1992}, and were important stepping stones towards the development of ion-trap based quantum computing~\cite{Cirac1995,Steane1997}.  Larger scale mechanical objects, such as fabricated nanomechanical resonators, have only recently been cooled close to their quantum mechanical ground state of motion~\cite{Groeblacher2009a,Park2009,Schliesser2009,Rocheleau2010,OConnell2010,Teufel2011b,Chan2011,Verhagen2011}.  In a pioneering experiment by O'Connell, et al.~\cite{OConnell2010}, a piezoelectric nanomechanical resonator has been cryogenically cooled ($T_b \sim 25$~mK) to its vibrational ground state and strongly coupled to a superconducting circuit qubit allowing for quantum state preparation and read-out of the mechanics.  An alternate line of research has been pursued in circuit and cavity optomechanics~\cite{ref:Braginsky1977}, where the position of a mechanical oscillator is coupled to the frequency of a high-$Q$ electromagnetic resonance allowing for back-action cooling~\cite{Wilson-Rae2007,Marquardt2007} and continuous position read-out of the oscillator. Such optomechanical resonators have long been pursued as quantum-limited sensors of weak classical forces~\cite{ref:Braginsky1977,Caves1980a,Regal2008,Schliesser2009,Clerk2010}, with more recent studies exploring optomechanical systems as quantum optical memories and amplifiers~\cite{Chang2011,Safavi-Naeini2011,Brooks2011,Massel2011}, quantum nonlinear dynamical elements~\cite{Ludwig2008}, and quantum interfaces in hybrid quantum systems~\cite{Wallquist2009,Stannigel2010,Safavi-Naeini2011a,Camerer2011}. 

Despite the major advances in circuit and cavity optomechanical systems made in the last few years, all experiments to date involving the cooling of mesoscopic mechanical oscillators have relied on careful measurement and calibration of the motionally scattered light to obtain the average phonon occupancy of the oscillator, $\nbar$.  Approach towards the quantum ground state in such experiments is manifest only as a weaker measured signal, with no evident demarcation between the classical and quantum regimes of the oscillator.  A crucial aspect of zero-point fluctuations of the quantum ground state, is that they cannot supply energy, but can only contribute to processes where energy is absorbed by the mechanics.  This is completely different from classical noise.  Techniques that attempt to measure zero-point motion without being sensitive to this aspect (i.e., standard continuous linear position detection) can always be interpreted classically and described by some effective temperature. 

A more direct method of thermometry and characterization of the quantized nature of a mechanical oscillator, one particularly suited to small $\nbar$ and utilized in the above-mentioned trapped atom experiments~\cite{Diedrich1989,Jessen1992,Monroe1995}, is referred to as motional sideband spectroscopy.  This method relies on the fundamental asymmetry in the quantum processes of absorption from (proportional to $\nbar$) and emission into (proportional to $\nbar+1$) the mechanical oscillator of phonons.  In the case of atomic systems, this asymmetry can be measured in the motionally-generated Stokes and anti-Stokes sidebands in either the fluorescence or absorption spectrum of the atom.  The ratio of the Stokes to anti-Stokes motional sideband amplitudes ($(\nbar+1)/\nbar$) deviates significantly from unity as the quantum ground state is reached ($\nbar\rightarrow 0$), and provides a self-calibrated reference for the phonon occupancy.  In the present experiment we cool a nanomechanical resonator to near its quantum ground state, and measure the asymmetry in the motional sidebands utilizing a form of resolved sideband spectroscopy based upon the filtering properties of a high-$Q$ optical cavity with linewidth narrower than the mechanical frequency.

\begin{figure}[t]
\begin{center}
\includegraphics[width=8.9cm]{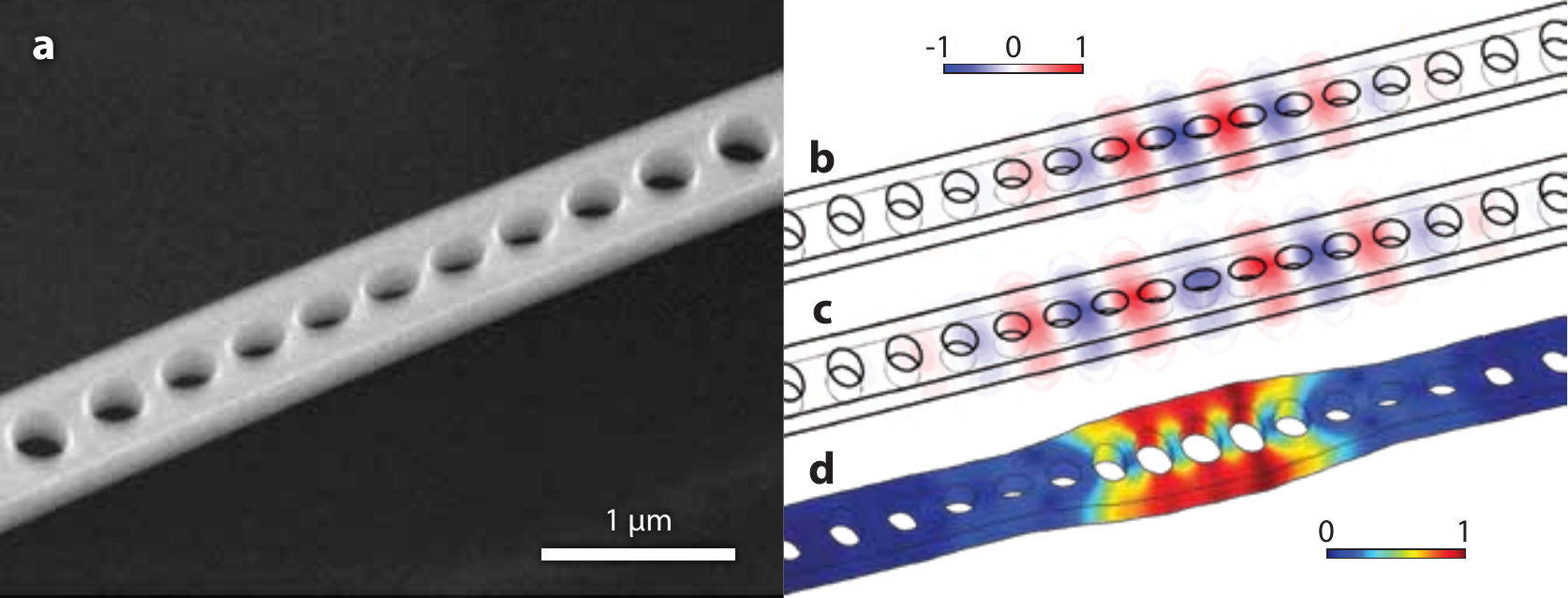}
\caption{\textbf{Multi-mode Optomechanical resonator.} \textbf{a}, A scanning electron micrograph of the silicon nanobeam optomechanical cavity.  \textbf{b}, and  \textbf{c}, show finite-element-method (FEM) numerical simulations of the first- and second-order optical modes of the cavity which are used for cooling and probing the mechanical motion, respectively. \textbf{d}, FEM numerical simulation of the coupled breathing mechanical mode.} \label{fig:OM_device}
\end{center}
\end{figure}

%% DESCRIBE THE SYSTEM
The cavity optomechanical system studied in this work consists of a patterned silicon nanobeam which forms an optomechanical crystal (OMC)~\cite{Eichenfield2009b} capable of localizing both optical and acoustic waves (see Fig.~\ref{fig:OM_device}).  The cavity is designed to have two optical resonances, one for cooling and one for read-out of mechanical motion.  The cooling mode is chosen as the fundamental mode of the patterned nanobeam cavity, with a frequency  $\omega_c/2\pi = 205.3~\text{THz}$ and a corresponding free-space wavelength of $\lambda_c=1460$~nm.  The read-out mode is the second-order mode of the cavity with $\omega_r/2\pi = 194.1~\text{THz}$ ($\lambda_r=1545$~nm). An in-plane mechanical breathing mode at $\omega_m/2\pi = 3.99~\text{GHz}$, confined at the center of the nanobeam due to acoustic Bragg reflection, couples to both optical resonances with zero-point optomechanical coupling rates $g_c/2\pi = 960~\text{kHz}$ and $g_r/2\pi = 430~\text{kHz}$ for the cooling and read-out modes, respectively ($g_c$ and $g_r$ are measured for each cavity mode from optically-induced damping of the mechanical mode as in Ref.~\cite{Chan2011}). The Hamiltonian of the coupled system is given by
\bea
\op{H}{}&= &\hbar \left(\omega_r + g_r \op{x}{}/x_\text{zpf}\right) \opdagger{a}{}\op{a}{} + \hbar \left(\omega_c + g_c \op{x}{}/x_\text{zpf}\right) \opdagger{c}{}\op{c}{}\nonumber\\
&&~~~~ + \hbar \omega_m \opdagger{b}{}\op{b}{},
\eea
where we have defined $\op{c}{}$ ($\opdagger{c}{}$) and $\op{a}{}$ ($\opdagger{a}{}$) to be the annihilation (creation) operators for photons in the cooling and read-out modes of the optical cavity. The optical modes are coupled to the mechanical displacement operator of the breathing mode, $\op{x}{} \equiv x_\text{zpf} (\opdagger{b}{} + \op{b}{})$, where $\opdagger{b}{}$ and $\op{b}{}$ are the creation and annihilation operators for phonons in the mechanical resonator, and $x_\text{zpf}$ is the zero-point fluctuation amplitude of the mechanical motion.

\begin{figure*}[t!!]
\begin{center}
\includegraphics[width=18.3cm]{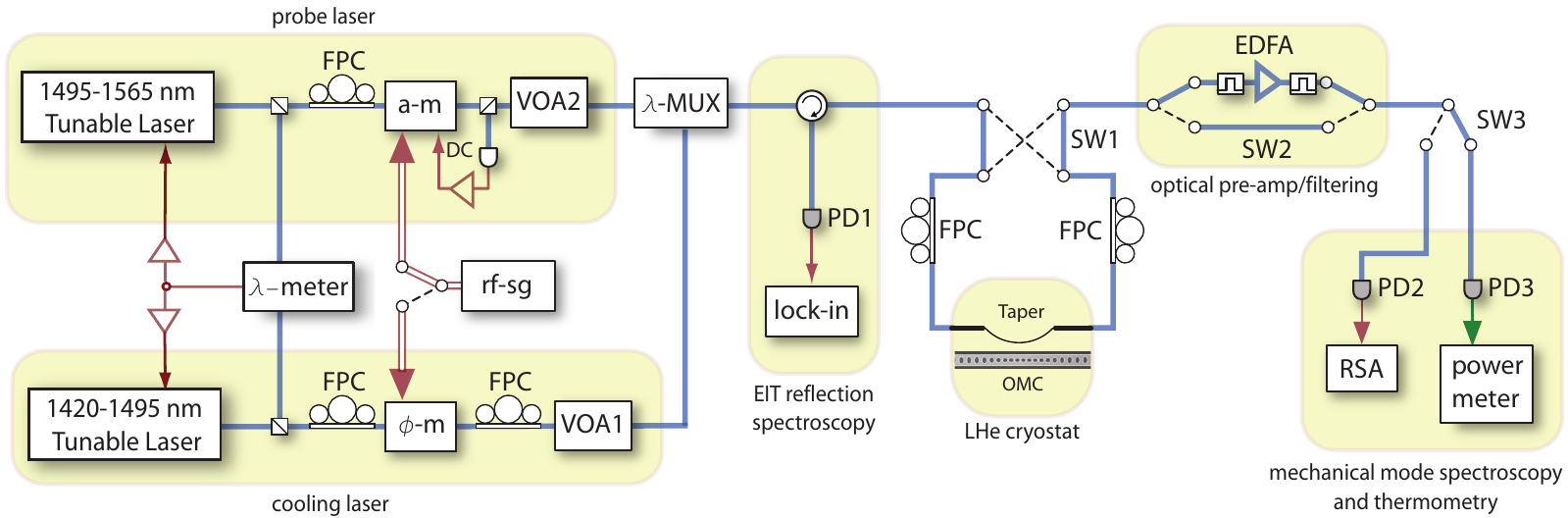}
\caption{\textbf{Experimental setup}. The silicon OMC cavity is mounted into a continuous flow Helium cryostat and optically coupled via an optical fiber taper probe.  Two narrowband lasers (New Focus Velocity series; linewidth $\sim 300$~kHz) are used to independently cool and read out the motion of the breathing mechanical mode of the OMC cavity.  The $1500$~nm (read-out) and $1400$~nm (cooling) laser beams are passed through variable optical attenuators (VOAs) to set the laser power, and combined at an optical fiber wavelength multiplexer ($\lambda-$MUX) before being sent into the cryostat through an optical fiber. Transmission of the $1500$~nm read-out beam through the OMC cavity, collected at the output end of the optical fiber, is filtered from the $1400$~nm cooling beam light via an optical fiber bandpass filter, pre-amplified by an Erbium-doped fiber amplifier (EDFA), and detected on a high-speed photodetector (PD2) connected to a real-time spectrum analyzer (RSA).  Other components are labeled as amplitude-modulation (a-m) and phase-modulation ($\phi$-m) electro-optic modulators, fiber polarization controller (FPC), swept frequency radio-frequency signal generator (rf-sg), lock-in amplifier (lock-in), optical switches (SW), and calibrated optical power meter (power meter).  Further details of the experimental apparatus and measurements are in the Methods section.}
\label{fig:setup}
\end{center}
\end{figure*}

An illustration of the experimental apparatus used to cool and measure the nanomechanical oscillator is shown in Fig.~\ref{fig:setup}.  In order to pre-cool the oscillator, the silicon sample is mounted inside a Helium flow cryostat. For a cold finger temperature of $6.3$~K, the bath temperature of the mechanical mode is measured to be $18~\text{K}$ (corresponding to a thermal phonon occupation of $n_b = 94$ phonons) through calibrated optomechanical thermometry as described in the Methods section and Ref.\cite{Chan2011}. At this temperature the mechanical damping rate to the thermal bath is found to be $\gamma_i/2\pi = 43~\text{kHz}$, corresponding to a intrinsic quality factor $Q_m = 9.2\times10^4$. The optical resonances of the OMC cavity are measured to have total damping rates of $\kappa_c/2\pi = 390~\text{MHz}$ and $\kappa_r/2\pi = 1.0~\text{GHz}$ for the cooling and read-out modes, respectively.  An optical fiber taper, formed from standard single mode optical fiber, is used to optically probe the OMC cavity via evanescent coupling.  The component of extrinsic damping resulting from coupling to the optical fiber taper waveguide is measured to be $\kappa_{e,c} /2\pi =46~\text{MHz}$ for the cooling mode and $\kappa_{e,r} /2\pi = 300~\text{MHz}$ for the read-out mode.

%% DESCRIBE COOLING PHYSICS

As alluded to above, resolved sideband cooling in optomechanical cavities follows physics which is formally similar to the Raman processes used to cool ions to their motional ground state~\cite{Diedrich1989}. A cooling laser, with frequency $\omega_l = \omega_c - \omega_m$, is tuned a mechanical frequency red of the cooling cavity resonance of the OMC, giving rise to an intra-cavity photon population $n_c$ at frequency $\omega_l$. The oscillations of the mechanical system cause scattering of the intra-cavity cooling beam laser light into Stokes and anti-Stokes sidebands at $\omega_c - 2\omega_m$ and $\omega_c$, respectively. Since the anti-Stokes sideband is resonant with the cavity at $\omega_c$, and $\kappa_c<\omega_m$, the anti-Stokes optical up-conversion process is greatly enhanced relative to the Stokes down-conversion process. This leads to cooling of the mechanical mode and can be modeled effectively as an additional mechanical damping term of $\gamma_\text{c} = 4 g_c^2 n_c / \kappa_c$ due to coupling to a near-zero temperature bath represented by the cooling laser.  Assuming a deeply resolved sideband system ($\kappa_c/\omega_m \ll 1$), the back-action cooled mechanical mode occupancy is approximately given by $\nbar_c=\gamma_in_b/(\gamma_i+\gamma_c)$~\cite{Wilson-Rae2007,Marquardt2007}.

%% DESCRIBE THERMOMETERY

\begin{figure}[t]
\begin{center}
\includegraphics[width=8.9cm]{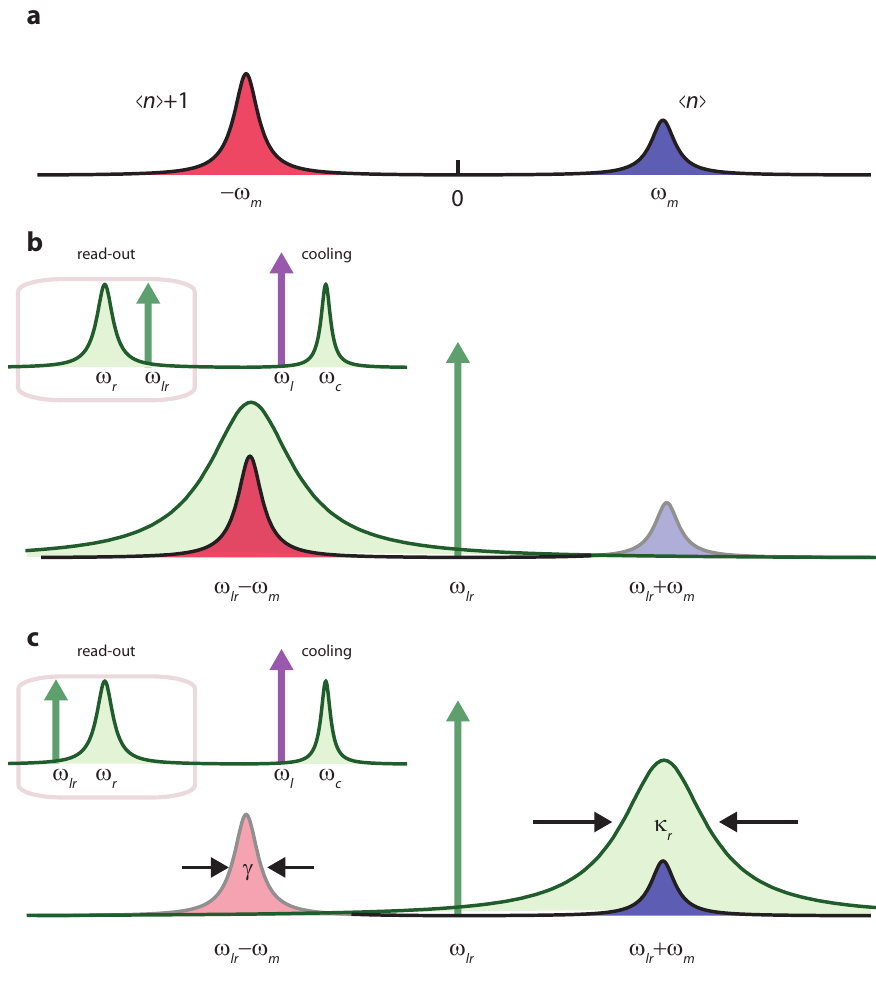}
\caption{\textbf{Optical filtering of motional sidebands}. \textbf{a}, Displacement noise PSD, $S_{xx}$, of a quantum simple harmonic
oscillator, plotted against $-\omega$ for clarity. \textbf{b}, Scheme for measurement of the down-converted (Stokes) motional sideband.  Here the read-out laser (green vertical arrow; frequency $\omega_{lr}$) is detuned a mechanical frequency above that of the read-out cavity resonance (green solid curve).  \textbf{c}, Corresponding scheme for measurement of the up-converted (anti-Stokes) motional sideband.  The linewidth of the readout cavity ($\kappa_{r}$) and the mechanical resonance ($\gamma$) are indicated.  Insets to \textbf{b} and \textbf{c} show a zoomed-out spectra indicating the relative frequency of the cooling cavity mode and cooling laser.}
\label{fig:spectra}
\end{center}
\end{figure}

Optical scattering of the intra-cavity light field can also be used to read out the motion of the coupled mechanical oscillator.  For a quantum harmonic oscillator, the autocorrelation $G_{xx}(t) \equiv \avg{\op{x}{}(t) \op{x}{}(0)}$ of the position~\cite{Clerk2010} is given by,
\be
G_{xx}(t)/x^2_\text{zpf} =  \nbar e^{+i\omega_m t - \gamma |t|}+(\nbar +1)e^{-i\omega_m t -\gamma |t|},
\ee
($\gamma$ is the total mechanical damping rate) with its Fourier transform, the displacement power spectral density (PSD), equal to
\bea
S_{xx}(\omega)/x^2_\text{zpf} &=& \frac{\gamma \nbar}{(\omega_m+\omega)^2+(\gamma/2)^2}\nonumber\\&&~~~ + \frac{\gamma (\nbar+1)}{(\omega_m-\omega)^2+(\gamma/2)^2},
\eea
The complex character of $G_{xx}(t)$ and the resulting asymmetry in frequency of $S_{xx}(\omega)$ (illustrated in Fig.~\ref{fig:spectra}a) arise from the non-commutativity of position ($\op{x}{}$) and momentum ($\op{p}{}$) operators in quantum mechanics and the resulting zero-point fluctuations.  This absorption-emission asymmetry has absolutely no classical analogue. Of course, at high phonon occupation numbers where $\nbar \approx \nbar+1$, the classically symmetric spectral density is recovered.  Since the optical cavity frequency is linearly coupled to the position of the mechanical oscillator, the displacement noise spectrum is imprinted on the photons leaving the cavity and can be measured optically, with the Stokes (anti-Stokes) motional sideband proportional to the positive (negative) frequencies of $S_{xx}(\omega)$.       

%Recent back-action cooling experiments~\cite{Schliesser2009,Rocheleau2010,Riviere2010,Teufel2011b,Chan2011} have simply measured the beating of the motionally scattered photons with the input laser light in the transmitted cooling beam.  In the resolved sideband regime, this produces a signal at the mechanical frequency proportional to $\nbar$, as the Stokes motional sideband (proportional to $\nbar + 1$) is suppressed by the cavity resonance.  As mentioned previously, absolute calibration of $\nbar$ from the measured cooling beam beat signal requires absolute calibration of the optical signals and their detection.  A measurement of both motional sidebands can be made by sending in laser light resonant with a second read-out optical cavity mode (so as neither motional sideband is preferentially scattered by the cavity), and heterodyning the transmitted signal with a frequency shifted component of the read-out laser, separating the beat notes of the Stokes and anti-Stokes sidebands.  Alternatively, as is done here, one can make a pair of measurements on the blue and red side of the read-out cavity in order to separately measure the two motional sidebands.  Details of this measurement and analysis are now given.

Consider a read-out laser with frequency $\omega_{lr}$ and detuning $\Delta \equiv \omega_r- \omega_{lr}$ from the read-out cavity mode.  The optical power spectrum about $\omega_{lr}$ of the transmitted read-out beam leaving the cavity is given by~\cite{Wilson-Rae2007},
%
%%%% !!!!!!!!!!!!!!!FACTORS!!!!!!!!!!!!
%
\bea
S(\omega) &=& |E_{r,\text{out}}|^2 \delta(\omega) +\frac{\kappa_{e,r}}{2\pi\kappa_r} \frac{A^{(r)}_-\gamma \nbar}{(\omega_m-\omega)^2+(\gamma/2)^2}\nonumber\\&&~~~ + \frac{\kappa_{e,r}}{2\pi\kappa_r} \frac{A^{(r)}_+\gamma (\nbar+1)}{(\omega_m+\omega)^2+(\gamma/2)^2},\label{eqn:Somega}
\eea
Here $A^{(r)}_+$ and $A^{(r)}_-$ are the detuning-dependent anti-Stokes and Stokes motional scattering rates, respectively, of the read-out laser, given by
\be
A^{(r)}_\pm = \frac{g_r^2 \kappa_r n_r}{(\Delta \pm \omega_m)^2+(\kappa_r/2)^2}.
\ee
As illustrated in Fig.~\ref{fig:spectra}b and c, the optical read-out cavity can be used to selectively filter the positive or negative frequency components of $S(\omega)$. For a detuning $\Delta = -\omega_m$ for the read-out laser, $A^{(r)}_+ \gg A^{(r)}_-$, resulting in a Lorentzian signal with area $I_{-}$ proportional to $\nbar+1$ centered at the mechanical frequency in the PSD of the photocurrent generated by the transmitted read-out laser.  Conversely, a detuning of $\Delta = \omega_m$ results in $A^{(r)}_- \gg A^{(r)}_+$, producing a signal of area $I_{+}$ proportional to $\nbar$.  Comparison of the area under the Lorentzian part of the measured photocurrent PSD of the transmitted read-out laser for detunings $\Delta = \pm \omega_m$, can then be used to directly infer the mechanical mode occupancy,
\be
\eta \equiv I_-/I_+ - 1 = \frac{1}{\nbar}\label{eqn:Iratio_ideal}.
\ee

This simple argument neglects the back-action of the read-out beam on the mechanical oscillator. In particular, the mechanical damping rate becomes detuning dependent, with $\gamma_\pm \equiv (\gamma_i + \gamma_\text{c})(1 \pm C_r)$ for $\Delta = \pm \omega_m$.  Here $C_r \equiv |A^{(r)}_+ - A^{(r)}_-|/(\gamma_i + \gamma_\text{c})$ is the effective cooperativity of the read-out beam in the presence of the strong cooling beam, and is found from the measured spectra by the relation
\begin{equation}
C_r  = \frac{\gamma_+ - \gamma_-}{\gamma_+ + \gamma_-}.\label{eqn:Cr}
\end{equation} 
The back-action of the read-out beam also results in a corresponding change in the phonon occupancy with read-out beam detuning, given by $\nbar_\pm =  \nbar_c/(1\pm C_r)$ for $\Delta = \pm \omega_m$, where $\nbar_c$ is the mechanical mode occupancy in the presence of only the cooling beam.  In our measurements this dependence is made small by ensuring that the read-out beam is much weaker than the cooling beam, so that $C_r \ll 1$.  Adding in a correction for the read-out laser back-action, one finds the following relation between the measured motional sidebands and the phonon occupancy of the cooled mechanical oscillator, 
\be
\eta^{\prime} \equiv \frac{I_-/I_+}{1+C_r} - \frac{1}{1-C_r} = \frac{1}{\nbar_c},\label{eqn:Iratio}
\ee
where for $C_r \ll 1$ we recover the standard relation given in Equation~(\ref{eqn:Iratio_ideal}).  

%\begin{figure*}[t!!]
%\begin{center}
%\includegraphics[width=1.85\columnwidth]{money_plot_vertical.pdf}
%\caption{\textbf{Measurement of Sideband Asymmetry}.  }
%\label{fig:asymmetry}
%\end{center}
%\end{figure*}

\begin{figure*}[t!!]
\begin{center}
\includegraphics[width=18.3cm]{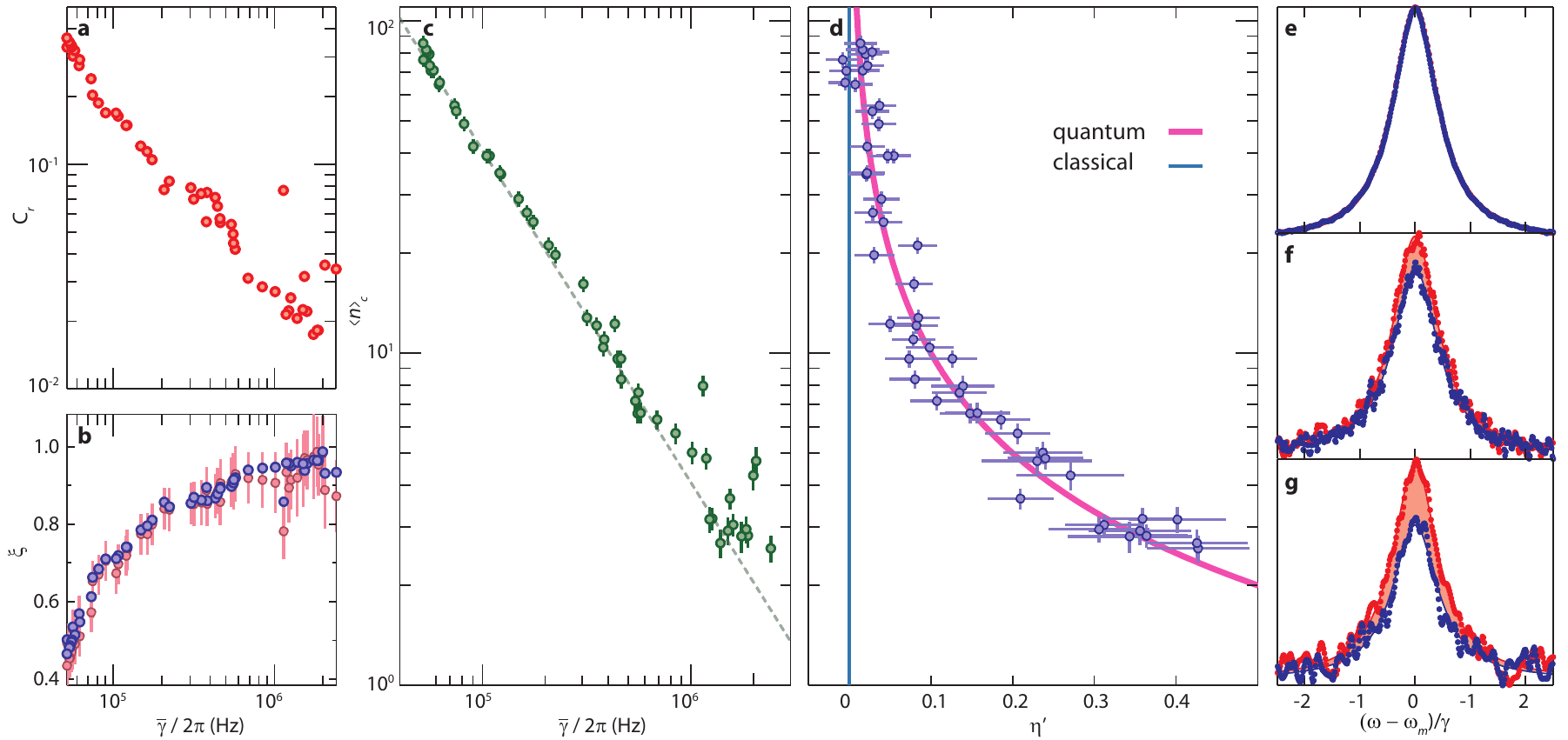}
\caption{\textbf{Measurement of motional sideband asymmetry}. \textbf{a} Plot of the cooperativity of the read-out beam as a function of damped mechanical linewidth. \textbf{b} Plot of the measured ratios $\gamma_-/\gamma_+$ (blue $\circ$) and $\nbar_+/\nbar_-$ (pink $\circ$).  \textbf{c} Plot of the laser cooled breathing mode phonon occupancy, $\nbar_c$, as a function of the optically damped mechanical linewidth, $\bar{\gamma}$. The dashed line is the predicted phonon number $\gamma_i n_b/\bar{\gamma}$ from an ideal back-action cooling model in which the thermal bath occupancy ($n_b$) and the intrinsic mechanical damping rate ($\gamma_i$) are constant and independent of the cooling beam power.  In practice both $n_b$ and $\gamma_i$ can vary with cooling beam power due to optical absorption in the silicon nanobeam cavity~\cite{Chan2011}, which accounts for some of the deviation of the measured phonon occupancy from the ideal cooling curve.  Vertical error bars in \textbf{b} and \textbf{c} indicate uncertainty in the calibrated phonon occupancy due to uncertainty in the system parameters and a 95\% confidence interval on the area from the Lorentzian fits.  \textbf{d} Plot of the asymmetry ($\eta^{\prime}$) in the measured Stokes and anti-Stokes sidebands of the read-out laser for each calibrated measurement of $\nbar_c$.  The horizontal error bars arise from a $2$\% uncertainty in the transmitted read-out laser beam power between detunings $\Delta=\pm\omega_m$, and a 95\% confidence interval in the Lorentzian fitting parameters of the measured spectra from which $I_{\pm}$ are calculated. The vertical error bars corresponding to uncertainty in $\nbar_c$ are the same as those plotted in \textbf{c}. The classical (blue curve) and quantum mechanical (pink curve) relations for the sideband asymmetry, given by $0$ and $1/\nbar_c$, respectively, are also plotted. \textbf{e-g}, Plot of the measured Stokes (red curve) and anti-Stokes (blue curve) read-out beam spectra for (from top to bottom) $\nbar_c = 85$, $6.3$, and $3.2$ phonons. For clarity, we have divided out the read-out back-action from each spectra by multiplying the measured spectra at detunings $\Delta = \pm\omega_m$ by $\gamma_\pm$ such that the ratio of the areas of the Stokes and anti-Stokes spectra are given by $I_-\gamma_-/I_+\gamma_+$, which approaches unity in the classical case. Additionally, we have plotted the horizontal axis in units of $\gamma$, and rescaled the vertical axis for different $\nbar_c$ to keep the areas directly comparable in each plot. The difference in the Stokes and anti-Stokes spectra, which arises due to the quantum zero-point fluctuation of the mechanical system, is shown as a shaded region in each of the plots.}
\label{fig:asymmetry}
\end{center}
\end{figure*}

%With the cooling laser locked to the red sideband of the fundamental cooling cavity mode, a series of measurements are performed versus cooling laser power

%% DESCRIBE PHONON THERMOMETRY MEASUREMENT WITH READ-OUT BEAM AND ROLE OF COOLING BEAM

Figure~\ref{fig:asymmetry} summarizes the measurement results of the calibrated mechanical mode thermometry and motional sideband asymmetry for the silicon nanobeam OMC cavity.  These measurements were performed with the cooling laser locked a mechanical frequency to the red of the fundamental mode of the OMC cavity, with the cooling laser power swept from $n_{c} \sim 1$ to $800$.  For each cooling laser power, the read-out laser is used to both estimate the mechanical mode phonon occupancy and to compare the motional sideband amplitudes. With the read-out laser set to a detuning from the read-out cavity mode of $\Delta = \omega_m$, a Lorentzian spectrum with linewidth $\gamma_+$ and integrated area $I_{+}$ is measured in the read-out laser photocurrent PSD, from which a mode occupancy of $\nbar_+$ is inferred from independently calibrated system parameters (see Methods and Ref.\cite{Chan2011}). Similarly, by placing the the read-out laser at $\Delta = -\omega_m$ we obtain spectra with linewidth $\gamma_-$ and integrated area $I_- \propto \nbar_-+1$, from which we estimate $\nbar_-$.  EIT spectroscopy~\cite{Safavi-Naeini2011} of the read-out optical cavity mode is also performed, providing a second, and more accurate (given the better signal-to-noise of such measurements) estimate of the mechanical damping rates $\gamma_{\pm}$. 

From the measured mechanical damping rates, the read-out cooperativity $C_r$ is found using relation~(\ref{eqn:Cr}), and plotted in Figure~\ref{fig:asymmetry}a. The ratios $\xi \equiv \gamma_-/\gamma_+$ and $\nbar_+/\nbar_-$ are plotted in Figure~\ref{fig:asymmetry}b, and found to be approximately equal in accordance with theory. From $\nbar_\pm$, the estimated phonon occupation number in the absence of a read-out beam, $\nbar_c$, is calculated and plotted in Figure~\ref{fig:asymmetry}c versus the mechanical damping rate $\bar{\gamma}=(\gamma_{+}+\gamma_{-})/2$. As expected, $\nbar_c$ drops approximately linearly with $\bar{\gamma}$, reaching a minimum value of approximately $2.6\pm0.2$ phonons, limited in this case by available cooling laser power.  As mentioned, the read-out beam was kept low in power so as to minimize the back-action in comparison to that of the cooling beam.  In particular, for the largest cooling powers $C_{r} \sim 2-3$\% and $n_\pm$ are within 10\% of $\nbar_c$.  

In Figure~\ref{fig:asymmetry}d, the measured values of the expression $\eta^{\prime}$ are plotted versus the calibrated value of $\nbar_c$. Also plotted are the classical and quantum values of this expression, $0$ and $1/\nbar_c$, respectively. A clear divergence from the classical result of $\eta^{\prime} = 0$ is apparent, agreeing with the deviation due to quantum zero-point fluctuations of the mechanical oscillator. At the largest powers, we measure asymmetries in the motional sideband amplitudes on the order of $40\%$ in agreement with the inferred $\nbar_c = 2.6$ phonons from calibrated thermometry.  This deviation is directly apparent in the measured spectra, shown for $\nbar_c=85$, $6.3$, and $3.2$ phonons in Figure~\ref{fig:asymmetry}e-g, with the shaded region corresponding to the noise power contribution due to quantum zero-point flucuations.

%% FUTURE

While the quantum nature of a mechanical resonator will come as little surprise to most physicists, its observation through the zero-point motion is a significant step towards observing and controlling the quantum dynamics of mesoscopic mechanical systems. By demonstrating the fundamentally quantum behaviour of an engineered mechanical nanostructure, we have shown that realizable optomechanical systems have the sensitivity and environmental isolation required for such quantum mechanical investigations.

%As electro- and optomechanical systems approach and enter the quantum regime, instrinsically direct methods of thermometry based on sideband asymmetry will become more prevalent. In the future, by moving towards more efficient coupling and detection schemes, and by improving our cryogenic technique (thus increasing the thermal decoherence time $\tau_T = \hbar Q_m / k_B T_b$ significantly~\cite{Alegre2011}), we expect to perform experiments with mechanical systems prepared in truly quantum states. Such states can be engineered using either quantum optical techniques~\cite{Aspelmeyer2010b} or coupling to qubits~\cite{OConnell2010} forming hybrid microwave-optical quantum systems. These investigations thus lead to new opportunties for experimental quantum information in engineered devices. By demonstrating the fundamentally quantum behaviour of an engineered mechanical nanostructure consisting of billions of atoms, we have shown that realizable optomechanical systems have the senstivity and environmental isolation required for experimental quantum optomechanics.

\begin{footnotesize}

%\bibliography{Mirror}

\section{Methods}

\textbf{Calibrated Thermometry.}\small{  In this work, an estimate of the mean phonon occupation number $\nbar$ is accomplished using the read-out optical cavity following a calibration procedure outlined in Ref.~\cite{Chan2011}. To briefly summarize (see experimental set-up in Fig.~\ref{fig:setup}), a $1500$~nm band read-out laser is sent through an amplitude electro-optic modulator (EOM), and combined with the cooling beam at a fiber optic multiplexer ($\lambda-$MUX). FPCs are used to ensure correct polarization before the EOMs and the OMC device. Calibration of the detuning of the read-out (cooling) laser from the read-out (cooling) cavity is performed using a form of EIT-spectroscopy~\cite{Chan2011,Safavi-Naeini2011} in which the reflection of optical sidebands generated on the read-out laser via a swept frequency radio-frequency signal generator (rf-sg) are monitored on PD1. The intra-cavity photon population generated by either the cooling ($n_c$) or read-out ($n_r$) laser beam is inferred from measurements of the power at the input and output of the taper, combined with a measurement of the asymmetry in the losses of the taper before and after the OMC device.  A wavemeter ($\lambda-$meter) is used to calibrate the optical linewidths of each cavity mode and to lock the lasers to within a few MHz of a given frequency.  The optomechanical coupling rate of the read-out cavity, $g_r$, is determined from the measured mechanical linewidth versus $n_r$ as measured in the transmitted read-out laser photocurrent spectrum on the RSA.  Calibration of the mechanical mode occupancy from the read-out laser photocurrent spectrum also requires knowledge of the optical power to detector voltage conversion ratio of PD2 ($G_e$) and the amplification furnished by the EDFA ($G_\text{EDFA}$). $G_e$ is determined by using SW3 to measure the total read-out laser intensity on the calibrated optical power meter.  $G_\text{EDFA}$ is determined by using SW2 to measure the amplitude of a signal generated by the rf-sg, with and without the EDFA in line.}

\noindent\textbf{Acknowledgments} The authors would like to thank Aash Clerk, Markus Aspelmeyer, and Simon Gr\"oblacher for their valuable input at various stages in this experiment.  This work was supported by the DARPA/MTO ORCHID program through a grant from AFOSR, and the Kavli Nanoscience Institute at Caltech. JC and ASN gratefully acknowledge support from NSERC.   

\end{footnotesize}

% \newpage
% \noindent
% \textbf{Figure Legends}
% \\
% \\
% \noindent
% FIG.~\ref{fig:OM_device}: \textbf{Opto-mechanical device.} \textbf{a}, SEM of central device + FEM of mechanical mode + FEM optical mode. \textbf{b}, SEM of phononic cage + FEM of mechanical modes (+ FEM of optical mode). \textbf{c}, show SEM of complete device.
% \\
% \\
% FIG.~\ref{fig:setup}: \textbf{Experimental setup}. Show the setup as in Jaspers notebook, also include a small SEM of the central part of the device and show how the cooling is actually done -- i.e.\ sketch a cavity response (Lorentzian), show the red-detuned laser beam and its stokes and anti-stokes sidebands, possibly also the probe for the EIT.
% \\
% \\
% FIG.~\ref{fig:spectra}: \textbf{Mechanical and optical response}. \textbf{a}, Show mechanical noise-power spectra for a blue and red-detuned probe (the geometric mean gives the Q!), where the curve under the peak is a proportional to $n_{phonon}$. The actual noise floor is due to the optical noise from the EDFA, also show the electronic noise floor from the detector and a theoretical prediction for the shot-noise. \textbf{b}, Optical response with EIT dip, maybe with an inset that zooms into the dip.
% \\
% \\
% FIG.~\ref{fig:cooling}: \textbf{a}, Cooperativity vs.\ $n_{photon}$. \textbf{b}, $T_{fit}$ vs.\ $T_{cryo}$. \textbf{c}, $n_{phonon}$ vs.\ $n_{photon}$, both experimental and predicted from compression factor -- with two insets of actual mechanical spectra at no cooling and strong cooling.

\end{document}